\documentclass[fleqn,twoside,twocolumn,nofootinbib]{revtex4} 
\usepackage{ujp} 
\begin{document}
\title[Kinetics of thermal excitation of a molecule in a condensed medium]
{Kinetics of thermal excitation of a molecule in a condensed medium}%
\author{V.I.Teslenko}
\affiliation{Bogolyubov Institute for Theoretical Physics, Nat. Acad. of Sci. of Ukraine}
\address{14b, Metrolohichna Str., Kyiv 03143, Ukraine}
\email{vtes@bitp.kiev.ua}
\author{D.Y.Iatsenko}
\affiliation{Kyiv National Taras Shevchenko University}
\address{2, Prospekt Glushkova, Kyiv 03022, Ukraine}

\udk{}  \razd{\secix}

\setcounter{page}{1}%
\maketitle

\begin{abstract}
A microscopic model for thermal excitation of vibrational ground state of a molecule interacting with a condensed medium is developed. The master equation for evolution of occupancies of the vibrational levels is derived. The rate constant limiting the process of molecular thermal excitation is determined analytically. It is shown that while this quantity is independent of temperature at low temperatures coinciding with a rate speed limit for quantum transitions due to uncertainty principle, at high temperatures it rather linearly rises with temperature according to the Einstein's relation for adiabatic transitions.
\end{abstract}

\section*{Introduction}
Considering kinetic and dynamic processes in different atomic and molecular nanostructures, for example, in molecular impurity centers, metal or semiconductor nanoparticles, biological macromolecules etc., requires the knowledge of microscopic mechanisms for vibrational relaxation of the system at limiting stages of its evolution. In recent years this problem became very important, particularly in the context of invention and implications of superfast heat generation in the close vicinity of gold nanoparticles under impulsive irradiation [1, 2]. From the physics point of view, such process is analogous to instantaneous forming of a point source of heat energy for molecules [3] followed by their thermal excitation and subsequent transition to higher vibrational levels. Therefore, in the frame of microscopic approach one has to solve the problem of impact of thermodynamic fluctuations in condensed medium which is at certain temperature on the process of vibrational relaxation of molecule, that was in ground vibrational state initially.

Many attempts were made to developing microscopic models for kinetic description of vibrational relaxation in molecules [4-9]. However, attention was mainly paid to analysis of intra- and intermolecular anharmonicity [5, 6, 9, 10] and role of spectral factors [9-12] involved in fluctuation-dissipation relations [9-11, 13]. At the same time, some specific types of random motions, particularly thermal noise (that is nonlinear over the molecular degrees of freedom), do not obey fluctuation-dissipation relations in the general case [14]. To approach such types of motion, in this communication we use a concept of generalized open quantum system that interacts with a heat bath [15, 16]. Within this framework one can take into account both bilinear intermolecular anharmonicity (that, according to the fluctuation-dissipation theorem, causes one-phonon relaxation transitions between molecular vibrational levels) and thermal noise (that randomly shifts these levels). We construct the microscopic model for thermal excitation of molecular vibrational ground state and derive the master equation for level occupancies. Moreover, we determine the rate constant of thermal excitation analytically and show that, in the general case, this quantity exhibits simple temperature dependence $\sim \{2\omega_0+(2\omega_c/\pi)exp(\hbar\omega_c/2kT)[exp(\hbar\omega_c/kT)-1]^{-1}\}$, where $\hbar$ and $k$ denote the Planck and Boltzmann constants, respectively, while the frequencies of vibrations in given molecule $\omega_0$ and molecular environment $\omega_c$ are determined only by structural factors and do not depend on the temperature $T$.

\section{Hamiltonian of a system}
We define a ground vibrational state as the state with highest kinetic energy $K(q)=-\frac{\hbar^2}{2m}\frac{\partial^2}{\partial q^2}$ and lowest potential energy. The latter in harmonic approximation takes the form $U(q,\{Q\})=U_0+(1/2)\{\partial^2 U(q,\{Q\})/\partial q^2\}|_{q=q_0}(q-q_0)^2$, where $U_0\equiv U(q,\{Q\})|_{q=q_0}=0$. Here $q$ and $q_0$ denote the normal nuclear coordinate of a given molecule (for simplicity we consider only one such coordinate with reduced mass $m$) and its equilibrium value found from the condition $\{\frac{\partial}{\partial q}U(q,\{Q\})\}|_{q=q_0}=0$, respectively. Potential energy depends also on displacements of the normal nuclear coordinates of molecules in the environment $\{Q\}$ from their equilibrium positions $\{Q_0\}$, that is necessary for occurrence of relaxation transitions between molecular vibrational levels. The interaction part of expansion of potential energy over mutual displacements $(q-q_0)(\{Q\}-\{Q_0\})$ responsible for such transitions appears in the form $V_{rel}=\{\frac{\partial^2}{\partial q\partial \{Q\}}U(q,\{Q\})\}|_{q=q_0,\{Q\}=\{Q_0\}}(q-q_0)(\{Q\}-\{Q_0\})$ [5, 6]. In this expansion the term $\sim (q-q_0)^3$ is excluded, that may be valid when the center of molecular vibrations coincides with the center of inversion. Also in $V_{rel}$ there are absent terms, that do not make a key contribution to kinetics of relaxation transitions on the limiting stage of establishment of the final equilibrium, as well as terms, that can be taken into account by renormalization of the Hamiltonian of the environment $H(\{Q\})$ [6] (i.e., terms $\sim(q-q_0)^2(\{Q\}-\{Q_0\})$; $(q-q_0)^4$ and $(\{Q\}-\{Q_0\})$; $(q-q_0)(\{Q\}-\{Q_0\})^2$ respectively). Thus, in the representation of oscillator wave functions $|n\rangle$ (where $n=0,1,2,...$ is the number of the vibrational state) the respective Hamiltonian of entire system (molecule+environment+interaction) takes the form:
\begin{equation}\label{eq1}
H(t)=H_0(t)+H_T+V_{rel}
\end{equation}
In this expression
\begin{equation}\label{eq2}
H_0(t)=\sum_n \hbar\omega(t)(n+1/2)|n\rangle\langle n|
\end{equation}
represents the main Hamiltonian of a molecule under consideration which is a stochastic operator depending on the random realizations of molecular movement in the environment along quasiclassical nuclear coordinates $\{Q(t)\}$ involved in the expression for frequency of molecular vibrations
\begin{equation}\label{eq3}
[\omega(t)]^2\equiv[\omega({Q(t)})]^2=\frac{1}{m}\{\frac{\partial^2}{\partial q^2}U(q,{Q(t)})\}|_{q=q_0}
\end{equation}
Besides, in equation (\ref{eq1})
\begin{equation}\label{eq4}
H_T=\sum_\lambda\hbar\Omega_\lambda(b_\lambda^+b_\lambda+1/2)
\end{equation}
is the Hamiltonian of a condensed medium, that in the harmonic approximation can be represented as a heat bath having an infinite set of the normal vibrations (phonons) with frequencies $\Omega_\lambda$. Quantities $b_\lambda^+$ and $b_\lambda$ refer to the operators of creation and annihilation of the respective $\lambda$th mode.
The last term in (\ref{eq1})
\begin{equation}\label{eq5}
V_{rel}=\sum_{n\lambda}\hbar\chi_\lambda\sqrt{n+1}(|n\rangle\langle n+1|+|n+1\rangle\langle n|)(b_\lambda^++b_\lambda)
\end{equation}
represents the operator of relaxation transitions in the one-phonon approximation, that describes transitions between the nearest vibrational levels. The corresponding parameter of bilinear intermolecular anharmonicity
\begin{equation}\label{eq6}
\chi_\lambda=\sqrt{\xi_\lambda\varpi_0\Omega_\lambda}
\end{equation}
characterizes the dependence of relaxation transitions on both the frequencies of normal vibrations in a heat bath $\Omega_\lambda$ and the shifted frequency of molecular vibrations $\varpi_0\equiv\omega({Q(t)})|_{\{Q\}=\{Q_0\}}$. This frequency shift appears due to interaction of a molecule with a random mean field of thermal motions in the environment. The respective non-shifted frequency is $\omega_0\equiv\varpi_0-\sigma$, where $\omega_0^2=\frac{1}{m}\{\frac{\partial^2}{\partial q^2}U^{(0)}(q)\}|_{q=q_0}$ with $U^{(0)}(q)$ being the time-independent part of the potential energy. The latter can be found from the condition $\{\frac{\partial}{\partial\{Q\}}U(q,\{Q\})\}|_{q=q_0}=0$. The corresponding standard mean field deviation is $\hbar\sigma=\frac{\hbar}{\sqrt{m}}\{\sqrt{[\frac{\partial^2}{\partial q^2}U(q,\{Q(t)\})]|_{q=q_0,\{Q\}=\{Q_0\}}}-\sqrt{[\frac{\partial^2}{\partial q^2}U^{(0)}(q)]|_{q=q_0}}\}$, while $\xi_\lambda=\frac{1}{4}\frac{(\{\partial^2 U(q,\{Q\})/\partial q\partial Q_\lambda\}|_{q=q_0,\{Q\}=\{Q_0\}})^2}{\{\partial^2 U(q,\{Q\})/\partial q^2\}|_{q=q_0}\{\partial^2 H(\{Q\})/\partial Q_\lambda^2\}|_{\{Q\}=\{Q_0\}}}$ defines the so called parameter of nonadiabaticity. The latter couples the molecular vibrations with those of the environment (in the case of adiabatic coupling the molecular coordinate must coincide with one of the normal coordinates $q=Q_{\lambda'}$ and the respective parameter is thus $\xi_{\lambda'}=1$).

From the general form of system Hamiltonian (\ref{eq1}) one can see that a problem of thermal excitation of a molecule in a condensed medium is reduced to the problem of description of evolution of harmonic quantum system (\ref{eq2}) having the randomly alternated energy levels (\ref{eq3}). By energy conservation law (principle of microscopic equilibrium) molecule exhibits persistent exchange of phonons with a heat bath (\ref{eq4}) through the one-phonon mechanism (\ref{eq5}). Thus, for derivation of the master equation it is reasonable to use approach that has been recently developed in papers [15, 16] just for such types of open quantum systems.

\section{Master equation}

To perform an analysis of behavior of a molecule during the process of its thermal excitation the master equation for evolution of observable state occupancies $\gamma_n(t)$ is required. Since the frequency of transitions between the molecular energy levels is the stochastic quantity, observable state occupancies appear to be averaged over random realizations of shifts of the energy levels: $\gamma_n(t)=\langle\langle \Gamma_n(t) \rangle\rangle$, where $\langle\langle ... \rangle\rangle$ denotes the averaging over random shifts, while $\Gamma_n(t)=\langle n|\rho(t)|n\rangle$ represent the non-averaged state occupancies. The molecular nonequilibrium density matrix $\rho(t)=tr_T\rho_{S+T}(t)$ is determined as a trace over the thermal bath states ($tr_T$) on the nonequilibrium density matrix of the entire system $\rho_{S+T}(t)$. In turn, evolution of the entire system is governed by the Liouville equation:
\begin{equation}\label{eq7}
\dot{\rho}_{S+T}(t)=-i\mathcal{L}(t)\rho_{S+T}(t)
\end{equation}
where $\mathcal{L}(t)=(1/\hbar)[H(t),...]$ is the Liouville stochastic superoperator that acts in the space of both dynamic and stochastic variables of the entire system Hamiltonian $H(t)$ (\ref{eq1}). Expanding (\ref{eq7}) into diagonal and off-diagonal components and solving obtained equations for diagonal component of the molecular density matrix $\rho_d(t)\equiv(tr_T\rho_{S+T}(t))_{diag}=\sum_n\Gamma_n(t)|n\rangle\langle n|$ yields $\dot{\rho}_d(t)=-\int_0^tdt'tr_T[\hat{T}_d\mathcal{L}_VS(t,t')\mathcal{L}_V\hat{T}_d\rho_{S+T}(t')]$, where $\mathcal{L}_V\equiv (1/\hbar)[V_{rel},...]$ and $S(t,t')=e^{-i\int_{t'}^td\tau(1-\hat{T}_d)\mathcal{L}(\tau)}$ with $\hat{T}_d$ being the projection superoperator which translates any operator into its diagonal component.

In many practical cases it is common that a characteristic time of transition processes $\tau_{tr}$ significantly exceeds the characteristic time of thermal relaxation $\tau_T$. This allows one to use a factorization  $\hat{T}_d\rho_{S+T}(t)=\rho_d(t)\rho_T$, where $\rho_T=exp(-H_T/kT)/tr_Texp(-H_T/kT)$ is the bath equilibrium density matrix. Thus, in Born approximation for the Liouville equation one arrives to the following coarse-grained master equation:
\begin{equation}\label{eq11}
\begin{aligned}
\dot{\rho}_d(t)=&-\frac{1}{\hbar^2}\int_0^tdt'tr_T\{\hat{T}_d[V_{rel},U(t,t')[V_{rel},\rho_d(t')\rho_T]\times\\
&\times U^+(t,t')]\}
\end{aligned}
\end{equation}
where $U(t,t')=\hat{T}exp[-(i/\hbar)\int_{t'}^td\tau(H_0(t)+H_T)]$ and $\hat{T}$ is the Dayson's chronological operator.

Taking into account (\ref{eq11}) and using exact form of system Hamiltonian (\ref{eq2}),(\ref{eq4}) and (\ref{eq5}), one can derive the equation for non-averaged state occupancies $\Gamma_n(t)$:
\begin{equation}\label{eq12}
\begin{aligned}
\dot{\Gamma}_n(t)=&\int_0^tdt'\{\mathcal{G}_{n+1n}(t,t')\Gamma_{n+1}(t')+\mathcal{G}_{n-1n}(t,t')\Gamma_{n-1}(t')\\
&-[\mathcal{G}_{nn+1}(t,t')+\mathcal{G}_{nn-1}(t,t')]\Gamma_n(t')\}
\end{aligned}
\end{equation}
Here $\mathcal{G}_{nm}(t,t')=2Re\sum_\lambda\chi_\lambda^2[(n+1)\delta_{mn+1}+n\delta_{mn-1}]e^{i\varpi_0(m-n)(t-t')}R_\lambda(t-t')\mathcal{F}_{nm}(t,t')$ are the non-Markovian transition probability densities that exhibit stochastic behavior through the stochastic functionals $\mathcal{F}_{nm}(t,t')=exp\{i\int_{t'}^td\tau[(m-n)(\omega(\tau)-\varpi_0)]\}$, while $R_\lambda(\tau)=N(\Omega_\lambda)e^{i\Omega_\lambda\tau}+[N(\Omega_\lambda)+1]e^{-i\Omega_\lambda\tau}$ correspond to the correlation functions for one-phonon transitions between nearest vibrational levels with $N(\Omega)=[exp(\hbar\Omega/kT)-1]^{-1}$ being the Bose distribution function.

The main difficulty in solving of the integro-differential equation (\ref{eq12}) lies in a non-Markovian character of the respective integrands. Moreover, one has to make explicit averaging of the stochastic functionals $\Gamma_n(t)$ and $\mathcal{F}_{nm}(t,t')$. However, when considering one-phonon transitions in the second order of perturbation theory over relaxation interaction, then the non-Markoviaty can be neglected [15]. Using this fact and taking into account that characteristic time of stochastic processes $\tau_{stoch}$ is of the order of thermal relaxation time $\tau_T\sim\tau_{stoch}$ and consequently $\tau_{stoch}<<\tau_{tr}$, one can perform stochastic averaging of (\ref{eq12}) in the non-explicit form, that yields:
\begin{equation}\label{eq13}
\begin{aligned}
\frac{\partial}{\partial t}\gamma_n(t)=&(n+1)W_+\gamma_{n+1}(t)+nW_-\gamma_{n-1}(t)- \\
&-[nW_++(n+1)W_-]\gamma_n(t)
\end{aligned}
\end{equation}
The respective averaged probabilities of transitions from the first to the zero (+) and from the zero to the first (-) vibrational levels are
\begin{equation}\label{eq14}
W_\pm=2Re\sum_\lambda\xi_\lambda\varpi_0\Omega_\lambda\int_0^\infty d\tau e^{\pm i\varpi_0\tau}R_\lambda(\tau)F(\tau)
\end{equation}
where $F(\tau)=\langle\langle exp\{i\int_0^\tau d\tau'[\omega(\tau')-\varpi_0]\}\rangle\rangle$ is the averaged correlation function of stochastic shifts.

As one can see, a general problem reduces to the problem of determining the model for levels' stochastic alternation $\omega(t)$, subsequent calculation of the correlation function $F(\tau)$ and final evaluation of transition probabilities $W_{\pm}$. Note that in the case of thermal excitation, initially only one energy level (zero) is occupied: $\gamma_n(0)=\delta_{n0}$. Thus, evolution of both the mean vibrational energy $\langle E_{vib}(t)\rangle\equiv\sum_n\hbar\varpi_0\gamma_n(t)(n+1/2)$ and state occupancies $\gamma_n(t)$ are determined by the well known expressions [4,5]:
\begin{equation}\label{eq15}
\left\{\begin{aligned}
&\langle E_{vib}(t)\rangle=\hbar\varpi_0[N(\varpi_0)(1-e^{-\kappa t})+1/2]\\
&\gamma_n(t)=\frac{(e^{\hbar\varpi_0/kT}-1)(1-e^{-\kappa t})^n}{(e^{\hbar\varpi_0/kT}-e^{-\kappa t})^{n+1}}
\end{aligned}\right.
\end{equation}
In the above equations $\kappa\equiv W_+-W_-$ is the effective rate constant of thermal excitation which fully describes kinetics of relaxation processes. Therefore, the quantity $\kappa$ represents a major kinetic parameter of interest which has been sought as a general solution of the problem considered in the one-phonon approximation. It should be noted that state occupancies could also be represented in the equivalent form $\gamma_n(t)=[\langle\gamma_n(t)\rangle-1/2]^n[\langle\gamma_n(t)\rangle+1/2]^{-n-1}$, where $\langle\gamma_n(t)\rangle\equiv N(\varpi_0)(1-e^{-\kappa t})+1/2$ is the average mean occupancy vibrational number.

\section{Rate constant}
Let us use symmetrical dichotomous process $\alpha(t)$ as a model for stochastic shifts of the molecular frequency $\omega(t)$ (\ref{eq3}) from its mean value $\varpi_0$. During such process, the frequency exhibits jumps between two equiprobable values $\omega(t)-\varpi_0\equiv\alpha(t)=\pm\sigma$ at random times with mean frequency $\nu$. It can be shown [15] that in this case in (\ref{eq14}) $F(\tau)=(k_1e^{-k_2\tau}-k_2e^{-k_1\tau})/(k_1-k_2)$, where $k_{1,2}=\frac{1}{2}(\nu\pm\sqrt{\nu^2-4\sigma^2})$. Then, according to (\ref{eq15}), the rate constant can be represented in the following general form:
\begin{equation}\label{eq16}
\begin{aligned}
\kappa=&2\varpi_0\sigma^2\nu\sum_\lambda\xi_\lambda\Omega_\lambda\{\frac{1}{[(\Omega_\lambda-\varpi_0)^2-\sigma^2]^2+\nu^2(\Omega_\lambda-\varpi_0)^2}\\
&-\frac{1}{[(\Omega_\lambda+\varpi_0)^2-\sigma^2]^2+\nu^2(\Omega_\lambda+\varpi_0)^2}\}
\end{aligned}
\end{equation}
From (\ref{eq16}) one can see, that rate depends on dynamic ($\omega_0,\Omega_\lambda,\xi_\lambda$), shifted ($\varpi_0=\omega_0+\sigma$), and stochastic ($\sigma,\nu$) system parameters. The last additionally provide for rate temperature dependence, as $\sigma=\sigma(T)$ and $\nu=\nu(T)$. Therefore, to perform further analysis, one has to represent rate constant in the convenient form.

First of all, let us make summation over the normal modes $\lambda$ in (\ref{eq16}). Let the corresponding factor of non-adiabaticity, that characterizes the constant of intermolecular anharmonicity (\ref{eq6}), has a structure
\begin{equation}\label{eq17}
\xi_\lambda=2\Omega_\lambda I(\Omega_\lambda)
\end{equation}
where the function $I(\Omega_\lambda)=\eta/(\Omega_\lambda^2+\eta^2)$ determines a density of spectral distribution of frequencies $\Omega_\lambda$. In this distribution the width parameter $\eta$ plays a role of adiabatic width for heat bath phonon spectra (in (14) the adiabatic limit $\xi_\lambda\rightarrow 1$ arrives at $\Omega_\lambda\rightarrow\eta$). Concrete value of $\eta$ depends on the model chosen for adiabatic interaction in given system. If vibrations in molecule and medium merge into the unified set of vibrational states, then a heat bath with an infinite set of normal modes forming an almost continuous spectrum can be chosen as a model for the molecular environment. The continuity of such spectrum means that after the transformation of a sum over $\lambda$ into an integral over $\Omega$ and its subsequent calculation the value of $\eta$ must be turned to zero: $\eta\rightarrow +0$.

This approximation is usually adequate in homogeneous structures of similar molecular nature, for example, in crystals. Another situation could  occur in heterogeneous systems (disordered and amorphous structures, biological macromolecules etc.), especially at finite temperatures. In such systems adiabatic spectral widths $\eta=\eta_\lambda$ would have additional dependence on the density of distribution of normal modes ${\lambda}$. In particular, their values might coincide with frequencies of certain adiabatic (mechanical) modes $\eta_\lambda\simeq\Omega_\lambda$ and would be much smaller ($\eta_\lambda<<\sigma$) than the standard deviation for thermal fluctuations $\sigma$. On the other hand, values $\eta_\lambda$ should considerably differ from nonadiabatic (optical) modes $\eta_\lambda<<\sigma<<\Omega_\lambda$. But in all cases, the shifted frequency of molecular vibrations $\varpi_0=\omega_0+\sigma$, that linearly rises with mean random field $\sigma$, significantly exceeds the adiabatic width: $\varpi_0>>\eta_\lambda$. Thus, in the one-phonon approximation $\Omega_\lambda\simeq\varpi_0$ the values of $\eta_\lambda$ must be considered small for all ${\lambda}$.

Note that physically, in the case of uniform distribution of normal modes, a width of adiabaticity $\eta$ forms in fact the finest scale over the frequency axis $\Omega_\lambda$. This allows one to make in (13), (14) a transformation to the variables $\Omega_\lambda\equiv(\Delta\Omega)\lambda=\eta\lambda$, where the elementary frequency shifts $\Delta\Omega\equiv\Omega_{\lambda+1}-\Omega_\lambda=\eta$ will not depend on the number of vibrational mode $\lambda$. By so doing, one can set in (13) the sum over $\lambda$ infinite and, taking into account the continuity condition $\eta=d\Omega\rightarrow 0$, make a transformation from a sum of functions $f(\Omega_\lambda)$ to the integral of a function $f(\Omega)$ by the following rule $\sum_{\lambda=1}^\infty f(\Omega_\lambda)=\lim_{\eta\rightarrow 0}[\frac{1}{2\pi\eta}\int_0^\infty f(\Omega)d\Omega]$. In a result, with the account of (14) elementary calculations yield:
\begin{equation}\label{eq18}
\kappa=\frac{4}{\pi}(\omega_0+\sigma)\int_0^{1+\frac{\omega_0}{\sigma}}dz\frac{\nu}{\sigma}[(1-z^2)^2+\frac{\nu^2}{\sigma^2}z^2]^{-1}
\end{equation}

We see that, since the sum (13) does not generally depend on the Bose distribution function for molecular vibrations $N(\varpi_0)$, the very knowledge of particular non-adiabaticity parameter $\xi_\lambda$ (14) allows us to reduce rather complicated equation (13) to the simple integral (15), which only contains the unknown dimensionless stochastic factors $z=\Omega/\sigma$, $\nu/\sigma$, and $\varpi_0/\sigma$. As a second step, let us explore a concrete model for the temperature dependence of stochastic parameters. For this purpose we use thermodynamic approach that was originally proposed in [15, 16]. In this approach, standard deviation of an amplitude of dichotomous process $\sigma$ is linked to the value of mean square of energy fluctuation $\overline{\delta E^2}$ via relation: $\pi^2\hbar^2\sigma^2=\overline{\delta E^2}$. On the other hand, from the theory of thermodynamic fluctuations in canonical ensembles we know that the latter quantity is given by $\overline{\delta E^2}=kT^2(\partial\bar{E}/\partial T)$, where $\bar{E}=2\pi\hbar\nu$ is the mean energy of dichotomous fluctuations with frequency $\nu$. In the classical limit, mean energy $\bar{E}=\hbar\omega_c[N(\omega_c)+1/2]$ per separate degree of freedom must turn out to thermal energy $kT$, while in the quantum limit, the same quantity must correspond to energy of zero-point oscillations with some correlation frequency $\omega_c$. Therefore, independently of the molecular system analyzed, the stochastic fluctuation parameters can be determined by the following expressions
\begin{equation}\label{eq18a}
\left\{\begin{aligned}
\sigma&=(\omega_c/\pi)\sqrt{N(\omega_c)[N(\omega_c)+1]}\\
\nu&=(\omega_c/2\pi)[N(\omega_c)+1/2]
\end{aligned}\right.
\end{equation}

General relations (\ref{eq18}) and (\ref{eq18a}) allow one to examine important limiting cases. Thus, since a characteristic scale for stochastic values (\ref{eq18a}) is determined by a frequency $\omega_c$, in the case of sufficiently low temperatures ($\hbar\omega_c>>kT$), the upper limit of the integral (\ref{eq18}) can be turned to infinity. In this case $\int_0^{\infty}dz\frac{\nu}{\sigma}[(1-z^2)^2+\frac{\nu^2}{\sigma^2}z^2]^{-1}=\pi/2$, and
\begin{equation}\label{eq19}
\kappa=2(\omega_0+\sigma)=2\omega_0+\frac{2\omega_c}{\pi}e^{\hbar\omega_c/2kT}[e^{\hbar\omega_c/kT}-1]^{-1}
\end{equation}

Conversely, at sufficiently high temperatures ($\hbar\omega_c<<kT$) numerical calculation of the integral in (\ref{eq18}) yields
\begin{equation}\label{eq20}
\kappa=\frac{2}{\pi}((A+4)\omega_0+A\sigma)+O(\frac{\omega_0}{\sigma})
\end{equation}
where $A\simeq 2.10357$. Note, that the value of integral in (\ref{eq18}) rapidly tends to $\pi/2$ with increasing of $\hbar\omega_0/kT$. It is equal to $\pi/2$ up to fourth order even when $\hbar\omega_0/kT=1$. Therefore, practically for all temperatures of interest the corresponding thermal excitation rate constant $\kappa$ can be well described by quite simple analytic relation (\ref{eq19}).

It is worth noting that equation (\ref{eq18}) can be proved only by certain approximations, that simplify general character of thermal excitation. Particularly, we use both the model of one-phonon transitions between nearest energy levels and dichotomous mechanism of frequency alternation due to the interaction with random motions in the environment.

The first approximation is a direct consequence of bilinearity of the intermolecular anharmonicity operator over mutual deviations of the nuclei of molecule under consideration and molecules in the environment from equilibrium positions (\ref{eq5}). As it is known, an account of non-linear terms significantly complicates the problem by making a relaxation of the system essentially multi-phonon. In particular, transitions over the one vibrational level become possible, as well as transitions accompanied by simultaneous creating or annihilating of several phonons in the medium [5, 6]. However, taking into consideration such effects is important only at very high temperatures, when the vibrational structure of molecule becomes undefined in a quantum sense. This is also the case for molecules with internal anharmonicity, when only the lowest vibrational levels are well defined, while energetically higher levels merge into a continuous spectrum.

The second approximation is not critical. Thus, using in (\ref{eq14}) instead of dichotomous model a common Gaussian model with analogous in their meaning stochastic parameters $\sigma$ and $\nu$, one achieves rate constant in the form
\begin{equation}\label{eq21}
\kappa=\frac{4}{\pi}(\omega_0+\sigma)\int_0^{1+\omega_0/\sigma}dz\frac{\nu}{\sigma}[1+\frac{\nu^2}{\sigma^2}z^2]^{-1}
\end{equation}
This equation due to the $\delta$-like character of its integrand can exactly be reduced to the same form (\ref{eq19}) as in the low temperature limit of dichotomous model (\ref{eq18}).

In general, the presence of two terms in expressions (15), (17), (18) for the rate constant $\kappa$ physically means, that in the harmonic (one-phonon) approximation the process of molecular thermal excitation is governed by the two additive mechanisms. The first is a quantum mechanism ($\sim 2\omega_0$) being essentially temperature-independent, while the second is quasiclassical one ($\sim 2\sigma$) being activated with temperature. The mutual contribution of these mechanisms is determined by characteristic frequencies of vibrations in the molecule ($\omega_0$) and medium ($\omega_c$) with relation to thermal frequency $\omega_T\equiv kT/\hbar$. Thus, in the quantum limit $\omega_0>\omega_c>>\omega_T$, the inequality $\omega_0>>\sigma$ holds, and the main contribution to thermal excitation is introduced by quantum transitions between nearest vibrational levels of the molecule. In this case $\kappa=2E_0/\hbar$ (17), that in a full accordance with the uncertainty principle corresponds to the rate speed limit for quantum transitions between orthogonal states with energy dispersion $E_0=\hbar\omega_0$ [17, 18]. In the limit of classical transitions $\omega_c<\omega_0<<\sigma<\omega_T$ and $\kappa=2\omega_T/\pi=2kT/\pi\hbar$, that coincides with the Einshtein relation for adiabatic rate of the Brownian oscillator relaxation [13, 14]. In this case, if holding the condition $\omega_c<\omega_0$, then according to adiabatic theorem there are no transitions between the energy levels of the molecule. As a result, thermal excitation occurs via simple increase of the shifted vibration frequency $\varpi_0=\omega_0+\sigma\approx\sigma=kT/\pi\hbar$ due to its direct coupling to random motions in the environment.

We have to note, that quantity $\kappa$ depends on the Planck constant $\hbar$ only through the characteristic time of thermal relaxation $\tau_T=\omega_T^{-1}=\hbar/kT$ involved in the Bose distribution function for correlation fluctuations in the medium $N(\omega_c)=[exp(\omega_c\tau_T)-1]^{-1}$. This property allows one to make a safe transformation from the quantum ($T\rightarrow 0$) to quasiclassical ($\hbar\rightarrow 0$) limit in a heat bath, considering processes of thermal relaxation to be very fast $\omega_T>>\omega_0>\omega_c$ ($\tau_T\rightarrow 0$) or setting temperature relatively high $T\rightarrow\infty$. By turn, the corresponding transformation from the temperature-independent to linear in temperature asymptotics for $\kappa$ is well-reproduced in the unified manner. Therefore, one does not need neither knowledge of the molecular ($\omega_0$) or correlation ($\omega_c$) vibrational frequencies, nor any introduction of additional factors for "correction" of the temperature dependence of rate constant in the classical limit [11]. This characteristic feature solves the so called harmonic oscillator paradox [19]. Indeed, one may think, that in accordance with the quantum theory the overall probability of transitions between the nearest levels of oscillator, for example between the $n$th and the $(n-1)$th, should be proportional to the total width of these levels (i.e. to the factor $\hbar^2\chi_\lambda^2(2n-1)(2|n\rangle\langle n|+1)$) (5), (11). But while turning to the classical limit $n\rightarrow\infty$, that is achieved at high temperatures (when $\omega_0<<kT/\hbar\rightarrow\infty$ or $N(\omega_0)\sim\langle|n\rangle\langle n|\rangle\rightarrow\infty$), this factor diverges. Thus, a concept of classical harmonic oscillator as a dynamic system being at the extremely high temperature and having an infinite set of densely distributed levels makes no sense. However, from general equations (12) it follows, that an evolution of both an average vibrational energy and oscillator state occupancies on thermal excitation depends on the rate constant $\kappa$ which is determined not by the sum, but by the difference between level widths, and nor between the some $n$th and the $(n-1)$th levels, but only between the first and the zero levels at any temperature and any density of level distribution. Because of this, both the numbers of vibrational levels and the respective Bose functions for molecular vibrations $N(\omega_0)$ remove from explicit consideration introducing any contribution to the value of rate constant $\kappa$ (13), that limits the process of establishing a final thermal equilibrium in the molecule.

\begin{abstract}
\end{abstract}

\end{document}